\documentclass[11pt,floatfix,noshowpacs]{revtex4}
\usepackage[dvips]{graphicx,color}

\usepackage[caption=false]{subfig}

\usepackage[latin1]{inputenc}
\usepackage{amsmath,amssymb,amsfonts,amsthm,latexsym}
\usepackage[english]{babel}
\usepackage{float}
\usepackage{subfig}
\usepackage{slashed}

\begin{document}

\title{\sc\Large{Magnetic susceptibility of the QCD vacuum in a nonlocal SU(3) PNJL model}}

\author{V.P. Pagura$^{a}$, D. G\'omez Dumm$^{b,c}$ S. Noguera$^{a}$ and N.N.\ Scoccola$^{c,d,e}$}

\address{
$^{a}$ Departamento de F\'{\i}sica Te\'orica and
Instituto de  F\'{\i}sica Corpuscular, Universidad de Valencia-CSIC, E-46100
Burjassot (Valencia), Spain\\
$^{b}$ IFLP, CONICET $-$ Dpto.\ de F\'{\i}sica, Fac.\ de Cs.\ Exactas,
Universidad
Nacional de La Plata, C.C. 67, 1900 La Plata, Argentina,\\
$^{c}$ CONICET, Rivadavia 1917, 1033 Buenos Aires, Argentina \\
$^{d}$ Departamento de F\'{\i}sica Te\'orica, Comisi\'on Nacional de Energ\'{\i}a Atómica,
Av.Libertador 8250, 1429 Buenos Aires, Argentina \\
$^{e}$ Universidad Favaloro, Sol{\'\i}s 453, 1078 Buenos Aires, Argentina
}

\begin{abstract}
The magnetic susceptibility of the QCD vacuum is analyzed in the framework
of a nonlocal SU(3) Polyakov-Nambu-Jona-Lasinio model. Considering two
different model parametrizations, we estimate the values of the $u$ and
$s$-quark tensor coefficients and magnetic susceptibilities and then we
extend the analysis to finite temperature systems. Our numerical results are
compared to those obtained in other theoretical approaches and in lattice
QCD calculations.
\end{abstract}

\maketitle


\section{Introduction}

One of the most interesting features of quantum chromodynamics (QCD) is the
nontrivial structure of its vacuum. This is clearly reflected on the vacuum
expectation values of scalar quark condensates $\langle\bar \psi_f
\psi_f\rangle$, which do not vanish at zero temperature and density. Light
quark condensates are usually taken as order parameters related to the
spontaneous breakdown of chiral symmetry, which can be regarded as one of
the most important aspects of low energy strong interaction physics. Now, in
order to get a more profound knowledge of the QCD vacuum it is interesting
to study hadronic systems in the presence of external sources. In
particular, it is seen that a constant external electromagnetic field
induces the existence of other nonvanishing condensates, which describe the
response of the vacuum to the source. We will concentrate here in the vacuum
expectation value (VEV) of the tensor polarization operator $\langle\bar
\psi\, \sigma_{\mu\nu}\, \psi\rangle$, where $\sigma_{\mu\nu} =
i[\gamma_\mu,\gamma_\nu]/2$ is the relativistic spin operator. In general,
to leading order in the external field, for each quark flavor $f$ one has
\begin{equation}
\langle\bar \psi_f\, \sigma_{\mu\nu}\, \psi_f\rangle_{A} \ = \ q_f
F_{\mu\nu}  \, \tau_f \ ,
\label{coefdef}
\end{equation}
where $F_{\mu\nu}$ is the field strength tensor, $q_f$ is the quark electric
charge and $\tau_f$ is the so-called tensor coefficient. The subindex $A$
indicates that the VEV is taken in the presence of an external
electromagnetic field $A^\mu$. It is also usual to introduce the parameters
$\chi_f$, defined by
\begin{equation}
\tau_f = \chi_f \; \langle\bar \psi_f \psi_f\rangle \ .
\end{equation}
In the literature $\chi_f$ is frequently referred to as the magnetic
susceptibility of the quark condensate, for a quark of flavor $f$. However,
notice that it only constitutes the spin contribution to the total magnetic
susceptibility. The quantity $\chi_f$ was first introduced in the context of
QCD sum rules~\cite{Ioffe:1983ju}. Later it was noted that it is also
relevant for the analysis of the muon anomalous magnetic
moment~\cite{Czarnecki:2002nt} and for the description of several processes
involving real photons, such as dijet production~\cite{Braun:2002en} and
radiative
decays~\cite{Aliev:1999tq,Ball:2003fq,Colangelo:2005hv,Rohrwild:2007yt}.

Previous calculations of $\tau_f$ and/or $\chi_f$ for light and
strange quark flavors have been carried out using QCD sum
rules~\cite{Belyaev:1984ic,Balitsky:1985aq,Ball:2002ps}, in the
holographic approach~\cite{Bergman:2008sg,Gorsky:2009ma}, using
the operator product expansion in the instanton liquid model and
chiral effective
models~\cite{Kim:2004hd,Dorokhov:2005pg,Goeke:2007nc}, using
zero-modes of the Dirac operator~\cite{Ioffe:2009yi}, and in
low-energy models of QCD such as the quark-meson
model and the Nambu-Jona-Lasinio (NJL)
model~\cite{Frasca:2011zn}. In addition, results from three-flavor
lattice QCD (LQCD) simulations have become available
recently~\cite{Bali:2012jv}. These include not only estimates at
zero temperature but also at temperatures in the region of the
chiral crossover transition. This has motivated the corresponding
analysis carried out in Ref.~\cite{Nam:2013wja} within an
effective SU(2) chiral effective model. The aim of the present
work is to extend these studies further by considering the
so-called nonlocal Polyakov-Nambu-Jona-Lasinio (nlPNJL)
models~\cite{Blaschke:2007np,Contrera:2007wu,Hell:2008cc,Contrera:2009hk,Hell:2009by},
in which quarks move in a background color field and interact among
themselves
through covariant nonlocal chirally symmetric four-point
couplings. These approaches, which can be considered as
improvements over the (local) PNJL
model~\cite{Meisinger:1995ih,Fukushima:2003fw,Megias:2004hj,Ratti:2005jh,
Roessner:2006xn,Mukherjee:2006hq,Sasaki:2006ww}, offer a common
framework to study both the chiral restoration and deconfinement
transitions. In fact, the nonlocal character of the interactions
arises naturally in the context of several successful approaches
to low-energy quark dynamics~\cite{Schafer:1996wv,RW94}. Moreover,
the presence of nonlocal form factors leads to a momentum
dependence in light quark propagators that, under an appropriate
choice of parameters~\cite{Noguera:2008, Noguera:2005ej}, is shown to be
consistent with the corresponding LQCD
results~\cite{bowman,Parappilly:2005ei,Furui:2006ks}.

The article is organized as follows. The theoretical framework is
presented in Sect.~II: in Sect.~II.A we describe the model and
quote the analytical expression for the tensor coefficients
$\tau_f$, and this is extended to finite temperature in
Sect.~II.B, where Polyakov-loop potentials are introduced; then in
Sect.~II.C we discuss the model parametrizations to be considered.
Sect.~III is devoted to present our numerical results, which are
compared with those obtained in other theoretical schemes and in
LQCD calculations. Our conclusions are sketched in Sect.~IV. We
also include three appendices. In Appendix A we provide some
details about the calculation of mean field values in our model,
while in Appendix B we describe our regularization prescription
for the tensor coefficients. Finally, in Appendix C possible
alternative calculations of the tensor coefficients in the NJL
model are discussed.

\section{Formalism}

\subsection{Magnetic susceptibility in a SU(3)$_f$ nonlocal chiral quark model}

As stated, we consider here a three-flavor nonlocal
Nambu-Jona-Lasinio (nlNJL)
model~\cite{Scarpettini:2003fj,Carlomagno:2013ona}. We will work
in Euclidean space, where the corresponding action is given by
\begin{eqnarray}
\label{se}
S_E &=& \int d^4x \ \left\{ \bar\psi(x)(-i \slashed \partial +
\hat m)\psi(x)-\frac{G}{2}\left[
j_a^S(x)j_a^S(x)+j_a^P(x)j_a^P(x)+j^{\,r}(x)j^{\,r}(x)\right] \right.
\nonumber\\
    &&\left. -\,\frac{H}{4}\, A_{abc}\left[
j_a^S(x)j_b^S(x)j_c^S(x)-3j_a^S(x)j_b^P(x)j_c^P(x)\right]
\right\} \ .
\end{eqnarray}
Here $\psi(x)$ is the $N_f=3$ fermion triplet $\psi = (u\ d\
s)^T$, and $\hat m={\rm diag}(m_u,m_d,m_s)$ is the current quark
mass matrix. We consider the isospin symmetry limit, assuming
$m_u=m_d$. The model includes flavor mixing through the 't
Hooft-like term driven by the coupling constant $H$, in which the
SU(3) symmetric constants $A_{abc}$ are defined by
\begin{equation}
\label{aabc}
A_{abc} \ = \
\frac{1}{3!}\,\epsilon_{ijk}\,\epsilon_{mnl}(\lambda_a)_{im}\,
(\lambda_b)_{jn}\,(\lambda_c)_{kl} \ ,
\end{equation}
where $\lambda_a$, $a=0,\dots,8$, are
the standard eight Gell-Mann matrices plus
$\lambda_0=\sqrt{2/3}\,\mathbf{1}_{3\times 3}$.
The fermion currents in Eq.~(\ref{se}) are given by
\begin{eqnarray}
j_a^S(x) &=& \int d^4z\; \mathcal{G}(z)\,
\bar \psi\left(x+\frac{z}{2}\right)\lambda_a
\psi\left(x-\frac{z}{2}\right)
\ , \nonumber\\
j_a^P(x) &=& \int d^4z\; \mathcal{G}(z)\,
\bar \psi\left(x+\frac{z}{2}\right)i\lambda_a \gamma_5
\psi\left(x-\frac{z}{2}\right)
\ , \nonumber\\
j^{\,r}(x)   &=& \int d^4z\; \mathcal{F}(z)\, \bar\psi\left(x+\frac{z}{2}\right)
\frac{i \overleftrightarrow{\slashed \partial}}{2\kappa}
\psi\left(x-\frac{z}{2}\right)\ ,
\label{curr}
\end{eqnarray}
where $\mathcal{G}(z)$ and $\mathcal{F}(z)$ are covariant form
factors responsible for the nonlocal character of the
interactions. Notice that the relative weight of the interaction
term that includes the currents $j^{\,r}(x)$ is controlled by the
parameter $\kappa$. This coupling leads to quark wave function
renormalization (WFR).

In what follows we will work within the mean field approximation
(MFA). In momentum space, the effective quark propagators can be
expressed as
\begin{equation}
S_f(p) = \frac{Z(p)}{-\slashed p + M_f(p)} \ ,
\label{quarkp}
\end{equation}
where $f=u,d,s$ is the corresponding quark flavor, and $M_f(p)$
and $Z(p)$ stand for the (momentum dependent) effective mass and
WFR, respectively. These are given by
\begin{eqnarray}
M_f(p) &=& Z(p)\, \Big[ m_f\, +\, \bar \sigma_f\, g(p)\Big] \ ,
\nonumber\\
Z(p) &=& \rule{0cm}{.8cm} \left[ 1\,-\,\frac{\bar \zeta}{\kappa}\,f(p)\right]^{-1}\
,
\label{mz}
\end{eqnarray}
where the functions $g(p)$ and $f(p)$ are the Fourier transforms
of $\mathcal{G}(z)$ and $\mathcal{F}(z)$, while $\bar \sigma_f$
and $\bar \zeta$ are mean field values of scalar fields associated
with the currents in Eq.~(\ref{curr}), in a flavor basis. Details
of the procedure to obtain these quantities are given in Appendix
A.

Let us consider in this framework the tensor polarization
operator. In the presence of an electromagnetic field $A^\mu$, the
vacuum expectation value of this operator at the leading order in
$A^\mu$ is given by
\begin{equation}
\langle \bar \psi_f(x) \; \sigma_{\mu\nu} \; \psi_f(x) \rangle_A \ = \ -\,
q_f \int\frac{ d^4 p}{(2\pi)^4}\frac{ d^4 p\,'}{(2\pi)^4} \ e^{i(p'-p)\cdot x}
A^\alpha(p-p\,') \ \mbox{Tr}\left[ \sigma_{\mu\nu} \, S_f(p) \
\Gamma_{f\alpha}(p,p\,') \ S_f(p\,') \right]\ , \label{tensor}
\end{equation}
where $\Gamma_{f\alpha}(p,p\,')$ stands for the effective quark-photon
vertex, and the trace is taken over Dirac and color indices. If the external
magnetic field is spatially uniform, the electromagnetic field can be
written as $A^\mu(x) = (-1/2)F^{\mu\nu}x_\nu$, therefore in momentum space
one has
\begin{equation}
A^\alpha(p-p\,') \ = \ -\frac{i}{2} \, F^{\alpha\beta} \,
\frac{\partial}{\partial p\,'_\beta} \left[(2\pi)^4 \delta^{(4)} (p-p\,')
\right]\ .
\end{equation}

In order to determine the couplings of dressed quarks to the
electromagnetic field, one has to take into account that within
the present nlNJL model the inclusion of gauge interactions
implies not only a change in the kinetic terms in the Lagrangian
(through the usual covariant derivative) but also a parallel
transport of the fermion fields entering the nonlocal currents in
Eq.~(\ref{curr}). As discussed in Ref.~\cite{Noguera:2005ej}, the
effective quark-photon vertex can be written as
\begin{eqnarray}
\Gamma_f^\alpha(p,p\,') &=&
\frac{1}{2} \left[ \frac{1}{Z(p)} +  \frac{1}{Z(p\,')} \right]
\gamma^\alpha +
\frac{1}{2} \left[ \frac{1}{Z(p)} -  \frac{1}{Z(p\,')} \right]
\frac{(p+p\,')^\alpha}{p^2-p\,'^2}\,
(\slashed p + \slashed p\,') \nonumber \\[3mm]
& &
- \left[ \frac{M_f(p)}{Z(p)} - \frac{M_f(p\,')}{Z(p\,')} \right]
\frac{(p+p\,')^\alpha}{p^2-p\,'^2}
+ \nu^{(1)}_f (p,p\,') \ T_1^\alpha + \nu^{(2)} (p,p\,') \ T_2^\alpha
\ ,
\label{gammaef}
\end{eqnarray}
where
\begin{eqnarray}
\nu^{(1)}_f (p,p\,') &=& -\,\frac{1}{p\,'^{\,2} - p^2} \ \int_{-1}^1 d\lambda \ \lambda
\left[ \frac{d}{dp^2} \frac{M_f(p)}{Z(p)} \right]_{p = \bar p-\lambda k/2} \ ,\nonumber \\
\nu^{(2)} (p,p\,') &=& \frac{1}{p\,'^{\,2} - p^2} \ \int_{-1}^1 d\lambda \ \lambda
\left[ \frac{d}{dp^2} \frac{1}{Z(p)} \right]_{p = \bar p-\lambda k/2}\ , \nonumber \\
T_1^\alpha (p,p\,') & = & p^\alpha \, (p\,' \cdot k) - p\,'^\alpha \, (p \cdot k)
\ , \nonumber \\
T_2^\alpha (p,p\,') & = & T_1^\alpha (p,p\,')\,\frac{\slashed p +\slashed
p\,'}{2}\ ,
\label{defs}
\end{eqnarray}
with $\bar p = (p+p\,')/2$, $k=p\,'-p$. The functions $\nu^{(1)}_f
(p,p\,')$ and $\nu^{(2)}(p,p\,')$ arise from the parallel
transport of fermion fields, which involves an integral over
an arbitrary path~\cite{ripka}. The result in
Eqs.~(\ref{gammaef}-\ref{defs}) corresponds to the choice of a
straight line path.

Then, taking into account the definition in Eq.~(\ref{coefdef}),
from Eqs.~(\ref{tensor}-\ref{defs}) the tensor coefficient is
found to be given by
\begin{equation}
\tau_f \ = \ 4 N_c \int \frac{ d^4 p}{(2\pi)^4} \ Z(p) \
\frac{M_f(p) - p^2 M_f'(p) }{\left[ p^2 + M_f(p)^2 \right]^2} \ ,
\label{t0}
\end{equation}
where $M_f'\equiv dM_f/dp^2$. Notice that this result does not
depend on the functions $\nu^{(1)}_f (p,p\,')$ and
$\nu^{(2)}(p,p\,')$, i.e.~on the arbitrary path chosen for the
gauge transformation carried out on fermion fields. It is also
worth noticing that for finite current quark masses the integral
in Eq.~(\ref{t0}) is ultraviolet divergent, thus it has to be
regularized. This can be accomplished by subtracting the
corresponding value in the absence of interactions (see the
discussion in Appendix B).

\subsection{Extension to finite temperature}

We will extend the analysis of the SU(3)$_f$ nlNJL model
introduced in the previous section to a system at finite
temperature by using the standard Matsubara formalism. In
addition, in order to account for confinement effects, we will
include the coupling of fermions to the Polyakov loop (PL),
assuming that quarks move on a constant color background field
$\phi = i g\,\delta_{\mu 0}\, G^\mu_a \lambda^a/2$, where
$G^\mu_a$ are the SU(3) color gauge fields. We will work in the
so-called Polyakov gauge, in which the matrix $\phi$ is given a
diagonal representation $\phi = \phi_3 \lambda_3 + \phi_8
\lambda_8$, taking the traced Polyakov loop $\Phi=\frac{1}{3} {\rm
Tr}\, \exp( i \phi/T)$ as an order parameter of the
confinement/deconfinement transition. Since
---owing to the charge conjugation properties of the QCD
Lagrangian~\cite{Dumitru:2005ng}--- the mean field traced Polyakov
loop is expected to be a real quantity, and $\phi_3$ and $\phi_8$
are assumed to be real valued~\cite{Roessner:2006xn}, one has
$\phi_8 = 0$, $\Phi = [1+ 2 \cos(\phi_3/T)]/3$. In addition, we
include effective gauge field self-interactions through a
Polyakov-loop potential ${\cal U}\,[\Phi]$. The resulting scheme
is usually denoted as a nonlocal Polyakov-Nambu-Jona-Lasinio
(nlPNJL) model~\cite{Blaschke:2007np, Contrera:2007wu,
Contrera:2009hk,Hell:2008cc,Hell:2009by}.

Concerning the PL potential, its functional form is usually based
on properties of pure gauge QCD. In this work we consider three
alternative forms that have been proposed in the literature. One
possible ansatz is that based on the logarithmic expression of the
Haar measure associated with the SU(3) color group integration.
The corresponding potential is given by~\cite{Roessner:2006xn}
\begin{equation}
\frac{{\cal{U}}_{\rm log}(\Phi ,T)}{T^4} \ =
\ -\,\frac{1}{2}\, a(T)\,\Phi^2 \;+
\;b(T)\, \log\left(1 - 6\, \Phi^2 + 8\, \Phi^3
- 3\, \Phi^4 \right) \ ,
\label{ulog}
\end{equation}
where
\begin{equation}
a(T) = a_0 +a_1 \left(\dfrac{T_0}{T}\right) + a_2\left(\dfrac{T_0}{T}\right)^2
\ ,
\qquad
b(T) = b_3\left(\dfrac{T_0}{T}\right)^3 \ .
\label{log}
\end{equation}
The parameters in these equations can be fitted to pure gauge
lattice QCD calculations so as to properly reproduce the
corresponding equation of state and Polyakov loop behavior. This
leads to~\cite{Roessner:2006xn}
\begin{equation}
a_0 = 3.51\ ,\qquad a_1 = -2.47\ ,\qquad a_2 = 15.2\ ,\qquad b_3 = -1.75\ .
\end{equation}
The values of $a_i$ and $b_i$ are constrained by the condition of reaching
the Stefan-Boltzmann limit at $T \rightarrow \infty$ and by imposing the
presence of a first-order phase transition at $T_0$, which is a further
parameter of the model. In the absence of dynamical quarks, from lattice
calculations one expects a deconfinement temperature $T_0 = 270$~MeV.
However, it has been argued that in the presence of light dynamical quarks
this temperature scale should be adequately reduced to about 210 and 190~MeV
for the case of two and three flavors, respectively, with an uncertainty of
about 30~MeV~\cite{Schaefer:2007pw}.

Besides the logarithmic
form in Eq.~(\ref{ulog}), a widely used potential is that given by a
polynomial function based on a Ginzburg-Landau
ansatz~\cite{Ratti:2005jh,Scavenius:2002ru}:
\begin{eqnarray}
\frac{{\cal{U}}_{\rm poly}(\Phi ,T)}{T ^4} \ = \ -\,\frac{b_2(T)}{2}\, \Phi^2
-\,\frac{b_3}{3}\, \Phi^3 +\,\frac{b_4}{4}\, \Phi^4 \ ,
\label{upoly}
\end{eqnarray}
where
\begin{eqnarray}
b_2(T) = a_0 +a_1 \left(\dfrac{T_0}{T}\right) + a_2\left(\dfrac{T_0}{T}\right)^2
+ a_3\left(\dfrac{T_0}{T}\right)^3\ .
\label{pol}
\end{eqnarray}
Here the reference temperature $T_0$ plays the same role as in the
logarithmic potential in Eq.~(\ref{ulog}). Once again, the parameters can be
fitted to pure gauge lattice QCD results so as to reproduce the
corresponding equation of state and Polyakov loop behavior. Numerical values
can be found in Ref.~\cite{Ratti:2005jh}.

Finally, we consider the so-called ``improved'' Polyakov loop
potentials recently proposed in Ref.~\cite{Haas:2013qwp}, in which
the full QCD potential ${\cal{U}}_{\rm glue}$ is related to a
Yang-Mills potential ${\cal{U}}_{\rm YM}$:
\begin{equation}
\frac{{\cal{U}}_{\rm glue}(\Phi ,t_{\rm glue})}{T ^4} \ = \
\frac{{\cal{U}}_{\rm YM}[\Phi ,t_{\rm YM}(t_{\rm glue})]}{T_{\rm YM}^4}\ ,
\label{utchica}
\end{equation}
where
\begin{equation}
t_{\rm YM}(t_{\rm glue}) \ = \ 0.57\, t_{\rm glue} \ = \
0.57 \left(\frac{T - T_c^{\rm glue}}{T_c^{\rm glue}}\right) \ .
\label{tglue}
\end{equation}
The dependence of the Yang-Mills potential on the Polyakov loop
$\Phi$ and the temperature $T_{YM}$ is taken from an ansatz such
as those in Eq.~(\ref{ulog}) or (\ref{upoly}), while for $T_c^{\rm
glue}$ a preferred value of 210~MeV is
obtained~\cite{Haas:2013qwp}.

Once the form of the effective action is established, the vacuum
expectation value of the tensor polarization operator at finite
temperature can be obtained by following a similar procedure as
the one described in the previous subsection. One gets
\begin{equation}
\tau_f (T) \ = \ 4 T \sum_{n=-\infty}^{\infty} \sum_{c=r,g,b}
\int \frac{ d^3 p}{(2\pi)^3} \ Z(p_{nc}) \,
 \frac{M_f(p_{nc}) - p_{nc}^2 \, M_f'(p_{nc})}{\left[p_{nc}^2 + M(p_{nc})^2\right]^2}\ ,
\label{taut}
\end{equation}
where $p_{nc}^2 = [(2n+1)\pi T+\phi_c]^2 + \vec p^{\;2}$, and
$\phi_c$ is given by the relation
$\phi=\textrm{diag}(\phi_r,\phi_g,\phi_b)=\textrm{diag}(\phi_3,-\phi_3,0)$.
In general, as in case of the $T=0$ expression in Eq.~(\ref{t0}),
it is seen that the integral in Eq.~(\ref{taut}) is ultraviolet
divergent. We regularize it by subtracting the $T=0$ divergent
piece, which is equivalent to subtract a ``free'' contribution
obtained from Eq.~(\ref{taut}) in the limit $\bar\sigma_{u,s} =
\bar\zeta = 0$, and add this contribution written in a regularized
form. Details are given in Appendix B.

\subsection{Model parameters and form factors}

In order to fully specify the model under consideration we need to
fix the value of the five parameters it includes, namely the
current quark masses $m_{u,s}$ and the coupling constants $G$,
$H$, and $\kappa$. In addition, one has to specify the form
factors ${\cal F}(z)$ and ${\cal G}(z)$ entering the nonlocal
fermion currents [or, equivalently, the corresponding Fourier
transforms $f(p)$ and $g(p)$]. Given the form factor functions,
one can fix the model parameters so as to reproduce the observed
meson phenomenology. Here, following
Ref.~\cite{Carlomagno:2013ona}, we will consider two
parametrizations, corresponding to two different functional forms
for $f(p)$ and $g(p)$. The first one corresponds to the often used
exponential behaviors
\begin{equation}
g(p)= \exp\left(-p^{2}/\Lambda_{0}^{2}\right) \ ,
\qquad f(p)= \exp\left(-p^{2}/\Lambda_{1}^{2}\right)\ ,
\label{ff1}
\end{equation}
which guarantee a fast ultraviolet convergence of the loop
integrals. Note that the range of the nonlocality in each channel
is determined by the parameters $\Lambda_0$ and $\Lambda_1$, which
can be viewed as effective momentum cutoffs. In order to fix the
parameters we have required the model to reproduce the
phenomenological values of five physical quantities, namely the
masses of the pseudoscalar mesons $\pi$, $K$ and $\eta^{\prime}$,
the pion weak decay constant $f_{\pi}$ and the light quark
condensate $\langle \bar\psi_u\psi_u \rangle$. In addition, on the
basis of lattice QCD estimations~\cite{Parappilly:2005ei}, we have
fixed the value of the quark WFR at zero momentum to be $Z(0) =
0.7$.

The second parametrization considered here is based on the
analysis in Ref.~\cite{bowman}, in which the effective mass
$M_u(p)$ is written as
\begin{equation}
M_u(p) \ = \ m_u \, + \, \alpha_m\, f_m(p)\ ,
\label{fm}
\end{equation}
where
\begin{equation}
f_m(p) \ = \ \frac{1}{1 + (p^2/\Lambda_0^2)^{3/2}}\ .
\end{equation}
From Eqs.~(\ref{mz}) one has $\alpha_m = (m_u\bar\zeta/\kappa
+\bar\sigma_u)/(1-\bar\zeta/\kappa)$. For the wave function renormalization
we use the parametrization~\cite{Noguera:2005ej,Noguera:2008}
\begin{equation}
Z(p) \ = \ 1 \, - \, \alpha_z\, f_z(p)\ ,
\label{fz}
\end{equation}
where
\begin{equation}
f_z(p) \ = \ \frac{1}{\left(1 + p^2/\Lambda_1^2\right)^{5/2}}\ .
\label{fzz}
\end{equation}
Here the new parameter $\alpha_z$ is given by $\alpha_z =
-\bar\zeta/(\kappa-\bar\zeta)$. The functions $f(p)$ and $g(p)$
can be now easily obtained from Eqs.~(\ref{mz}), (\ref{fm}) and
(\ref{fz}). As shown in
Refs.~\cite{Carlomagno:2013ona,Noguera:2008}, for an adequate
choice of parameters these functional forms can reproduce very
well the momentum dependence of quark mass and WFR obtained in
lattice calculations. We complete the model parameter fixing by
taking as phenomenological inputs the values the of the pion, kaon
and $\eta'$ masses and the pion weak decay constant.

In Table~I we quote the numerical values for the
model parameters that we have obtained for the above-described
form factor functions. In what follows, the parametrizations
corresponding to Eqs.~(\ref{ff1}) and (\ref{fm}-\ref{fzz}) will be
referred to as parametrizations PI and PII, respectively.
\begin{table} [ht]
\begin{center}
\begin{tabular}{c c c}
\hline \hline
 & \ \ \ \ \ PI \ \ \ \ \ & \ \ \ \ PII \ \ \ \ \\
\hline \hline
$m_u$ [MeV] & 5.7  & 2.6 \\
$m_s$ [MeV] & 136  & 64.9 \\
$G\Lambda_0^2$  & 23.64  & 16.65  \\
$-H\Lambda_0^5\ \ $ & 526 & 202.8 \\
$\kappa$ [GeV] & 4.36 & 8.218 \\
$\Lambda_0$ [GeV] & 0.814  & 0.795 \\
$\Lambda_1$ [GeV] & 1.032  & 1.510  \\
\hline
\end{tabular}
\caption{\small{Model parameters for the form factors in Eqs.~(\ref{ff1})
(PI) and (\ref{fm}-\ref{fzz}) (PII).}}
\end{center}
\end{table}

\section{Numerical results}

Given the model parametrization we can solve the set of
equations~(\ref{gaps}) and (\ref{zeta}), which allow us to obtain the mean
field values $\bar \sigma_{u,s}$ and $\bar\zeta$ at zero temperature, as
described in App.~A. Once these values are obtained it is straightforward to
compute the quark condensates and the tensor coefficients for light and
strange quark flavors, according to Eqs.~(\ref{cond}) and (\ref{t0}). Our
numerical results for parametrizations I and II are summarized in
Table~\ref{tableII}, where we also quote for comparison the corresponding
estimates obtained within other models. Firstly we observe that our model,
in accordance with other theoretical results, predicts a diamagnetic
behavior for the QCD vacuum. In addition, for both parametrizations the
values obtained for the $u-$tensor coefficient are found to be in very good
agreement with the LQCD estimate. In the case of the light quark magnetic
susceptibility we find some discrepancy between the results for PI and PII,
which turn out to be above and below the LQCD estimate, respectively. The
discrepancy can be explained by noting that the values for the light quark
condensates for both parametrizations are also significantly different. In
fact, this difference arises basically from the fact that for PI we have
taken as input the phenomenological value $-\langle \bar
\psi_u\,\psi_u\rangle^{1/3}=240$~MeV, which corresponds to a renormalization
scale of about 1~GeV, while PII has been obtained through a fit to lattice
data in Ref.~\cite{Parappilly:2005ei} for the effective quark propagator,
which correspond to a higher momentum scale of 3~GeV. Regarding the
$s-$tensor coefficient, we find that its value is more dependent on the
chosen parametrization than in the case of $\tau_u$. When comparing with the
quoted results of other models, it is seen that the prediction obtained
from PII is the closest one to LQCD. In any case, it is important to point
out that in general the various theoretical scenarios leading to the results
presented in Table II consider different renormalization scales, therefore
the comparison of numerical values should be taken with some care. Moreover,
in the case of the predictions obtained within the local NJL model it is
seen that the results are rather dependent on the calculation scheme. This
is discussed in App.~C, where we compare the values for $\tau_u$ arising
from different regularization approaches. For comparison we include in
Table~\ref{tableII} the results given by the calculation in
Ref.~\cite{Frasca:2011zn} and those arising from an alternative approach
that we refer to as ``weak field propagator expansion'' (WFPE), in which we
have used the SU(3)$_f$ NJL parametrization given in
Ref.~\cite{Hatsuda:1994pi}.
\begin{table} [h]
\begin{center}
\begin{tabular}{ccccccccc}
\hline \hline
                   &        & \multicolumn{2}{c}{nlNJL}
                            & \hspace*{.2cm} NJL    \hspace*{.2cm}
                            & \hspace*{.2cm} NJL$^*$    \hspace*{.2cm}
                            & \hspace*{.2cm}  ILM   \hspace*{.2cm}
                            & \hspace*{.2cm}  DS    \hspace*{.2cm}
                            & \hspace*{.2cm} LQCD   \hspace*{.2cm}          \\
                            &           & \hspace*{.4cm} PI  \hspace*{.4cm} &  \hspace*{.4cm}  PII   \hspace*{.4cm}   &          &   &   &    &             \\
\hline \hline
$\mu$                       &   [GeV]      &  0.814 & 3.0          & 0.627    & 0.631 & 0.85 & 0.4-0.7  & 2.0            \\
$m_u$                       &   [MeV]      & 5.7    & 2.6          & 5        &  5.5  & 5    &  0      & 3.5            \\
$-\langle\bar \psi_u \, \psi_u\rangle^{1/3}$ &   [MeV]      & 240    & 316          & 253      &  247  & 260  &  251     & 269            \\
$-\langle\bar \psi_s \, \psi_s\rangle^{1/3}$ &   [MeV]      & 198    & 341          &  $...$        &  267  &  $...$      &    $...$      &  250  \\
$\tau_u$                    &   [MeV]    & 38.2   & 44.6         &  69      & 25.8  & 40-45  &  28-33   & 40             \\
$\tau_s$                    &   [MeV]    &  9.7   &  30          &  $...$        & 19.8  & 6-10   & $...$         & 53             \\
$-\chi_u\ \ $               & [GeV$^{-2}$] & 2.77   & 1.42         &  4.3     &  1.72 & 2.5    & 1.7 -2.1 & \hspace{.2cm} 2.05 (0.09)  \hspace{-.8cm}  \\
$-\chi_s\ \ $               & [GeV$^{-2}$] & 1.25   & 0.76         &   $...$       &  1.03 & $...$     &   $...$       & \hspace{.2cm} 3.40 (1.40) \hspace{-.8cm}    \\
\hline
\end{tabular}
\caption{Condensates and magnetic susceptibilities obtained in the present
nlNJL model in comparison with other existing theoretical estimates: NJL
corresponds to the NJL model calculation in Ref.~\cite{Frasca:2011zn},
NJL$^*$ corresponds to a NJL model calculation based on what we call WFPE
approach (see App.~C), ILM to the instanton liquid model calculation in
Ref.~\cite{Kim:2004hd}, DS to the Dyson-Schwinger calculation in
Ref.~\cite{Watson:2013ghq} and LQCD to the lattice estimate in
Ref.~\cite{Bali:2012jv}. }\label{tableII}
\end{center}
\end{table}

\begin{figure}[htb]
\includegraphics[width=0.50\textwidth]{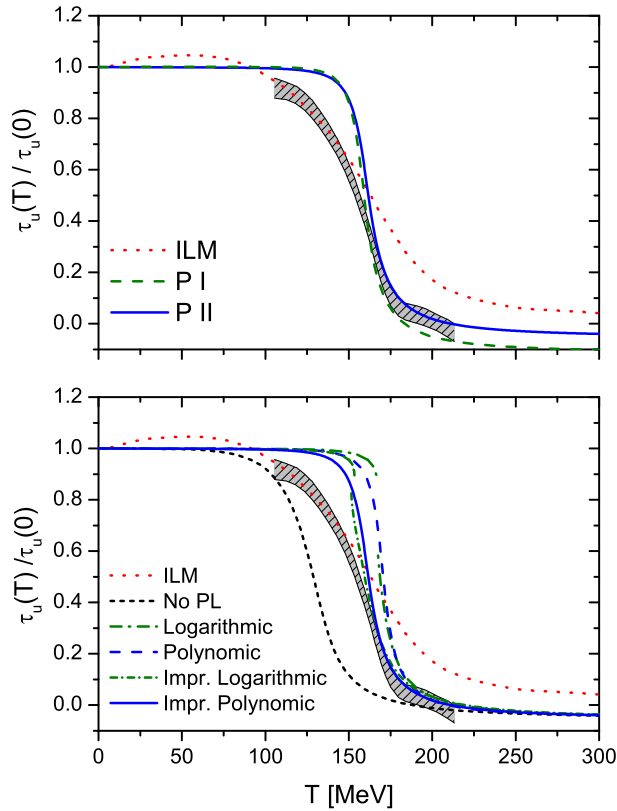}
\caption{Normalized $u$-quark tensor coefficient vs.\ temperature for
nonlocal PNJL models. Upper panel: results corresponding to the polynomial
PL potential in Eq.~(\ref{upoly}), for parametrizations PI and PII. Lower
panel: results corresponding to parametrization PII, for various PL
potentials. For comparison, values obtained within the
ILM~\cite{Nam:2013wja} (dotted lines) and results from
LQCD~\cite{Bali:2012jv} (dashed grey bands) are also shown in both graphs.}
\label{tau-T}
\end{figure}

Let us now study the temperature dependence of the tensor coefficient in the
various possible scenarios available for the parametrizations and Polyakov
loop potentials discussed in the previous section. Our results for $\tau_u$
as a function of the temperature, using the regularization prescription
discussed in App.~B, are presented in Fig.~\ref{tau-T}. If the temperature
is increased starting from $T=0$, for all cases under consideration it is
seen that the tensor coefficient remains approximately constant up to some
critical temperature, and then one finds a sudden drop, which is a signature
of the restoration of the SU(2) chiral symmetry. Therefore, $\tau_u$ may be
regarded as an approximate order parameter for the chiral restoration
transition. In the upper panel of Fig.~\ref{tau-T} we show the curves
obtained within the present nlPNJL model for both parametrizations PI and
PII (dashed and solid lines, respectively), considering the improved polynomial potential
for the Polyakov loop. For comparison we also show the results from
Ref.~\cite{Kim:2004hd}, obtained in the context of the instanton liquid
model (ILM), as well as LQCD estimates from Ref.~\cite{Bali:2012jv} (dotted
line and grey dashed band, respectively). Firstly we notice that ILM results
predict that for low temperatures the tensor condensate becomes increased
with respect to $\tau_u(0)$, while in our model it remains approximately
constant up to the chiral transition region. In order to characterize the
transition, we define the critical temperature $T_c$ as the temperature at
which the function $\tau_u(T)$ has an inflection point. Following this
definition we find for the case of the improved polynomial potential a
critical temperature $T_c =158$ (160)~MeV for PI (PII), while lattice
results lead to $T_c^{\rm LQCD}\sim 162$~MeV~\cite{Bali:2012jv}. Moreover,
we observe that at temperatures above the transition region the shape of the
curves obtained within our model are in reasonable agreement with lattice
calculations. On the other hand, the onset of the transition within nlPNJL
models is found to be rather steep, thus the curve arising from the ILM
seems to be more compatible with lattice results right below the critical
temperature. In fact, this discrepancy between nlPNJL and LQCD estimates may
be cured once the mesonic fluctuations are included in the Euclidean
action~\cite{Blaschke:2007np,Hell:2008cc, Hell:2009by,Radzhabov:2010dd}.
This can be understood by noting that when the temperature is increased the
light mesons should be excited before the quarks, and this would soften the
behavior of the tensor coefficient at the onset of the transition. It is
important to mention that the incorporation of mesonic corrections should
not modify the critical temperatures.

Next, in the lower panel of Fig.~\ref{tau-T} we show the curves for the
tensor coefficient as function of the temperature for the nlPNJL model,
considering various functional forms for the Polyakov loop potential. All
results correspond to the lattice QCD-inspired parametrization PII. For
comparison we include once again LQCD and ILM results, as well as a curve
corresponding to the tensor coefficient in the nlNJL model without the
coupling between the quarks and the Polyakov loop. In this last case
(short-dashed line) it is observed that the transition temperature turns out
to be too low in comparison with LQCD estimates, as it is indeed expected
from previous calculations~\cite{Contrera:2007wu}. The graph shows that
whereas different PL potentials give rise to different shapes for
$\tau_u(T)$ at temperatures below the chiral transition, once the transition
is surpassed the functions converge to a single curve that is in agreement
with lattice estimates. In general it is seen that for the polynomial
potentials the transition is smoother than in the case of the logarithmic
ones, for which the transition is found to be of first order. Furthermore,
for the improved potentials the curves tend to be smoother and the agreement
with lattice results starts to occur already at the transition temperature.
These general features on the comparison between the results of nlPNJL
models and LQCD have also been observed within the study of chiral
restoration, taking (as it is usually done) the quark condensates as order
parameters of the transition~\cite{Carlomagno:2013ona}.

Finally, let us briefly discuss our results for the $s$-quark tensor
coefficient and for other possible regularization prescriptions for
$\tau_u(T)$ and $\tau_s(T)$. As stated, the results in Fig.~\ref{tau-T}
correspond to the prescription introduced in App.~B, which is consistent
with the usual regularization carried out at zero temperature. However, it
may be argued that there are other possible regularization procedures. One
possible option is to define tensor coefficients $\tau_f^{(\rm int)}$ by
taking the expression in Eq.~(\ref{tautreg}) without the addition of the
free regularized terms, i.e.\ $\tau_f^{(\rm int)}(T)\equiv\tau_f^{(\rm
reg)}(T)-\tau_f^{(0,\rm reg)}(T)$. This means to keep just the contribution
of strong interaction dynamics to the tensor coefficients, hence in the
limit of large temperatures one gets $\tau_f^{(\rm int)}(T)\to 0$ instead of
the asymptotic free quark system behavior given by Eq.~(\ref{taularget}).
Furthermore, another way to get rid of the free quark contribution at high
temperatures is to define a ``subtracted tensor coefficient'' $\tau_{\rm
sub}$ as
\begin{equation}
\tau_{\rm sub}(T) \ = \ \tau_u^{(\rm reg)}(T) -
\frac{m_u}{m_s}\,\tau_s^{(\rm reg)}(T) +
\frac{m_u\,N_c}{2\pi^2}\,\log\bigg(\frac{m_s}{m_u}\bigg)\ ,
\end{equation}
which by construction also vanishes at large $T$ [see
Eq.~(\ref{taularget})].
\begin{figure}[hbt]
\includegraphics[width=0.50\textwidth]{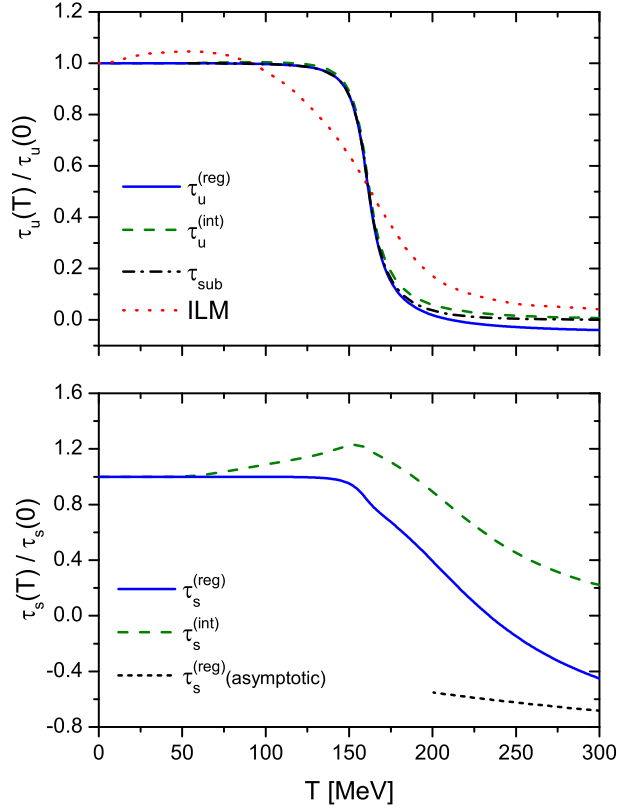}
\caption{Normalized tensor coefficients vs.\ temperature for nlPNJL models.
Results correspond to parametrization PII, improved polynomial PL potential
and different regularization prescriptions. Upper panel: results
corresponding to the $u$-quark tensor coefficients $\tau_u^{(\rm reg)}$ and
$\tau_u^{(\rm int)}$ and the subtracted tensor coefficient $\tau_{\rm sub}$.
For comparison values obtained within the ILM~\cite{Nam:2013wja} are also
shown. Lower panel: results corresponding to the $s$-quark tensor
coefficients $\tau_s^{(\rm reg)}$ and $\tau_s^{(\rm int)}$. The short dashed
line shows the asymptotic behavior of $\tau_s^{(\rm reg)}(T)/\tau_s^{(\rm
reg)}(0)$ at large temperatures. } \label{tau-sT}
\end{figure}

The curves corresponding to $\tau_u^{\rm (reg)}(T)$, $\tau_u^{\rm (int)}(T)$
and $\tau_{\rm sub}(T)$, together with ILM results, are shown in the upper
panel of Fig.~\ref{tau-sT}. In all cases the results are normalized to the
values at $T=0$, and correspond to parametrization PII and the improved PL
potential discussed in Sect.~II.B. For temperatures below the transition, it
is found that $\tau_u^{\rm (reg)}(T)$ and $\tau_{\rm sub}(T)$ keep constant,
while $\tau_u^{\rm (int)}$ (dashed line) shows some increase. This growth,
barely noticeable in the figure, is in any case negligible in comparison
with that found in the case of the ILM. Then, at the transition region the
shape of all three curves look very much alike, therefore it can be said
that the transition features do not depend on the regularization
prescription. At larger temperatures, as expected, the curves for
$\tau_u^{\rm (int)}(T)$ and $\tau_{\rm sub}(T)$ are similar (in both cases
the contribution of free quarks has been somehow excluded), while
$\tau_f^{\rm (reg)}$ is governed by the logarithmic behavior given by
Eq.~(\ref{taularget}). The curves corresponding to the $s$-quark tensor
coefficients $\tau_s^{\rm (reg)}(T)$ and $\tau_s^{\rm (int)}(T)$ are shown
in the lower panel of Fig.~\ref{tau-sT}. As in the case of the $u$ quark, it
is seen that $\tau_s^{\rm (reg)}$ remains approximately constant for low
$T$, starting to decrease at about the chiral transition critical
temperature. However, we find that the slope is not so pronounced as in the
case of $\tau_u$. At large temperatures the curve approaches the asymptotic
logarithmic behavior, shown by the short-dashed line in the figure. On the other
hand, the behavior of $\tau_s^{\rm (int)}(T)$ is quite different, showing a
significant increase at temperatures below the transition and then a
relatively slow descent. Thus, it is seen that even if the behavior of the
$s$-quark tensor coefficient reflects the chiral restoration, it cannot be
taken as a suitable order parameter in order to determine the critical
transition temperature.

\section{Summary and conclusions}

In this work we have investigated the magnetic susceptibility of the QCD
vacuum in the framework of a nonlocal SU(3) Polyakov-Nambu-Jona-Lasinio
model.

Firstly we have considered the situation at vanishing temperature. We have
found that the values for the $u$-quark tensor coefficient $\tau_u$ obtained
within our model for parameterizations I and II are similar to each other,
the results being in good agreement with estimates from lattice QCD and
instanton liquid model calculations. On the other hand, these values are
somewhat above the result obtained within a Dyson-Schwinger approach and
clearly below the value arising from the NJL model calculation of Ref.~[17].
It should be taken into account, however, that ---as discussed in App.~C---
NJL model results are quite dependent on the way in which the calculation is
performed. For the the corresponding $u$-quark magnetic susceptibilities we
find some discrepancy between the results arising from our parametrizations
I and II. This can be understood by noting that the values for the
respective chiral condensates are also different to each other, which, in
turn, is related to the fact that the parametrizations correspond to
different momentum scales. In the case of the $s$-quark quantities our
predictions turn out to be in general more dependent on the chosen
parametrization. It should be noticed that lattice QCD estimates are also
subject to larger uncertainties in this case.

Concerning the results at finite temperature we find that the tensor
coefficient $\tau_u$ remains approximately constant up to a critical
temperature, at which there is a sudden drop that can be clearly identified
with the restoration of the SU(2) chiral symmetry. The curves are found to
be similar for different regularization prescriptions. The
stability observed at low temperatures differs from the behavior predicted in
the context of the instanton liquid model, which shows a noticeable bump in
that region. As occurs for other quantities (e.g.~the scalar quark
condensates) in the framework of nlPNJL models at the mean field level, we
notice that at the onset of the chiral transition the behavior of the
tensor coefficient is rather steep in comparison with lattice QCD estimates.
This discrepancy is expected to be cured once meson fluctuations are
included in the calculation. In any case, these corrections should not
modify the behavior of the tensor coefficient above the transition, which is
found to be in good agreement with lattice QCD results.

\section*{Acknowledgments}

This work has been partially funded by CONICET (Argentina) under
Grants No.\ PIP 578 and PIP 449, by ANPCyT (Argentina) under
Grants No.\ PICT-2011-0113 and PICT-2014-0492, by the National
University of La Plata (Argentina), Project No.\ X718, by the
Mineco (Spain) under contract FPA2013-47443-C2-1-P and by the 
Centro de Excelencia Severo Ochoa Programme grant SEV-2014-0398 
and by Generalitat Valenciana (Spain) under contract
PrometeoII/2014/066.

\section*{Appendix A: Mean field approximation and gap equations at $T=0$}

\renewcommand{\theequation}{A.\arabic{equation}}
\setcounter{equation}{0}

Details on how to deal with the action in Eq.~(\ref{se}) at the
mean field level can be found e.g.~in
Ref.~\cite{Carlomagno:2013ona}. For the reader's convenience, in
this appendix we sketch just the main details. We start by
performing a standard bosonization of the fermionic theory,
introducing scalar fields $\sigma_a(x)$, $\zeta(x)$ and
pseudoscalar fields $\pi_a(x)$, together with auxiliary fields
${\cal S}_a(x)$, ${\cal R}(x)$ and ${\cal P}_a(x)$, with
$a=0,\dots , 8$. Now we follow the stationary phase approximation,
replacing the path integral over the auxiliary fields by the
corresponding argument evaluated at the minimizing values
$\tilde{\cal S}_a$, $\tilde{\cal R}$ and $\tilde{\cal P}_a$. Next,
we consider the MFA, in which the scalar and pseudoscalar fields
are expanded around their vacuum expectation values:
\begin{equation}
\label{mfaf}
\sigma_a(x) = \bar \sigma_a+\delta\sigma_a(x) \ , \qquad
\zeta(x) = \bar\zeta+\delta\zeta(x) \ , \qquad
\pi_a(x) = \delta\pi_a(x) \ .
\end{equation}
We have assumed that pseudoscalar mean field values vanish, owing
to parity conservation. Moreover, for the scalar fields only
$\bar\sigma_{0,8}$ and $\bar\zeta$ can be different from zero due
to charge and isospin symmetries. For the neutral fields
($a=0,3,8$) it is convenient to change to a flavor basis,
$\sigma_a,\pi_a \to \sigma_f,\pi_f$, where $f=u,d,s$, or
equivalently $f=1,2,3$. Then, the mean field action reads
\begin{eqnarray}
\label{semfa}
\frac{S_E^{\,\mbox{\tiny MFA}}}{V^{(4)}}
&=& 2\,N_c
\sum_{f} \int \dfrac{d^3p}{(2\pi)^3}\; \log
\left[\dfrac{Z(p)^2}{p^2 + M_f(p)^2} \right]
\nonumber \\
 & &  - \left(\bar\zeta\, \bar {\cal R} + \dfrac{G}{2}\,\bar {\cal R}^2 +
\dfrac{H}{4}\,\bar {\cal S}_u\, \bar {\cal S}_d \, \bar {\cal S}_s \right)
- \dfrac{1}{2}\, \sum_f \left( \bar \sigma_f \bar {\cal S}_f
+ \dfrac{G}{2}\, \bar {\cal S}_f^2 \right)\ ,
\end{eqnarray}
where $N_c$ is the number of colors, and $\bar {\cal S}_f$ and
$\bar {\cal R}$ stand for the values of $\tilde{\cal S}_f$ and
$\tilde{\cal R}$ within the MFA, respectively. The functions
$M_f(p)$ and $Z(p)$, given by Eqs.~(\ref{mz}), correspond to
the momentum-dependent effective masses and WFR of quark
propagators $S_f(p)$ in Eq.~(\ref{quarkp}).

By minimizing the mean field action in Eq.~(\ref{semfa}) one
gets the set of coupled gap equations
\begin{eqnarray}
\bar \sigma_u + G\bar {\cal S}_u+\frac{H}{2}\bar {\cal S}_d \bar {\cal S}_s &=& 0 \ , \nonumber\\
\bar \sigma_d + G\bar {\cal S}_d+\frac{H}{2}\bar {\cal S}_s \bar {\cal S}_u &=& 0 \ , \nonumber\\
\bar \sigma_s + G\bar {\cal S}_s+\frac{H}{2}\bar {\cal S}_u \bar {\cal S}_d &=& 0 \ ,
\label{gaps}
\end{eqnarray}
plus an extra equation arising from the $j^{\,r}(x)$ current-current
interaction,
\begin{equation}
\bar\zeta +G\bar {\cal R}  =  0 \ .
\label{zeta}
\end{equation}
The mean field values $\bar {\cal S}_f$ and $\bar {\cal R}$ in these equations
are given by
\begin{eqnarray}
\bar {\cal S}_f &=& -\,8N_c \int \frac{d^4p}{(2\pi)^4}\ g(p)\;
\frac{Z(p)\, M_f(p)}{p^2 + M_f(p)^2} \ \, , \qquad f=u,d,s \; , \nonumber\\
\bar {\cal R} &=& \frac{4 N_c}{\kappa} \int \frac{d^4p}{(2\pi)^4}\ p^2\, f(p)
\; \sum_{f=1}^3 \;\frac{Z(p)}{p^2 + M_f(p)^2} \ \, .
\label{sr}
\end{eqnarray}
Thus, for a given set of model parameters and form factors, from
Eqs.~(\ref{mz}) and (\ref{gaps}-\ref{sr}) one can numerically obtain the
mean field values $\bar \sigma_{f}$ and $\bar\zeta$.

The quark-antiquark condensates $\langle \bar \psi_f\, \psi_f
\rangle$ can be now easily calculated by taking the derivative of
the Euclidean mean field action with respect to the current quark
masses. One gets
\begin{equation}
\langle \bar \psi_f\, \psi_f \rangle \ = \ -\,4N_c \int
\frac{d^4p}{(2\pi)^4}\ \left[ \frac{M_f(p)}{p^2 + M_f(p)^2} -
\frac{m_f}{p^2 + m_f^2}\right]\ \, ,
\label{cond}
\end{equation}
where we have subtracted a ``free quark condensate'' in order to
regularize the otherwise divergent momentum integral.

\section*{Appendix B: Regularization of the tensor coefficient}

\renewcommand{\theequation}{B.\arabic{equation}}
\setcounter{equation}{0}

As in the case of the quark condensate, the expression for the
tensor coefficient in Eq.~(\ref{t0}) can be regularized by
subtracting a ``free'' $T=0$ contribution obtained in the limit
$\bar \sigma_{u,s} = \bar \zeta = 0$ (see
e.g.~Ref.~\cite{Kim:2004hd}):
\begin{equation}
\tau_f^{\rm (reg)} \ = \ 4 N_c \int \frac{ d^4 p}{(2\pi)^4} \ \bigg\{
Z(p) \
\frac{M_f(p) - p^2 M_f'(p) }{\left[ p^2 + M_f(p)^2 \right]^2} \ -
\frac{m_f}{\big(p^2 + m_f^2\big)^2} \bigg\}\ .
\label{t0reg}
\end{equation}
In the same way, for the case of a system at finite temperature
$T$ we regularize the divergent integral in Eq.~(\ref{taut}) by
subtracting a finite temperature contribution in which $\bar
\sigma_{u,s} = \bar \zeta = 0$. Then, in order to recover the
proper finite $T$ behavior at large $T$, we add this contribution after
subtracting the $T=0$ divergent piece as in Eq.~(\ref{t0reg}).
Thus, using the same definitions as in Eq.~(\ref{taut}), we have
\begin{eqnarray}
\tau_f^{\rm (reg)}(T) & = & 4 T \sum_{n=-\infty}^{\infty}
\sum_{c=r,g,b} \int \frac{ d^3 p}{(2\pi)^3} \ \bigg\{
Z(p_{nc}) \,
\frac{M_f(p_{nc}) - p_{nc}^2 \, M_f'(p_{nc})}{\left[p_{nc}^2 + M(p_{nc})^2\right]^2}
\nonumber \\
& & \ - \ \frac{m_f}{\big(p_{nc}^2 + m_f^2\big)^2}
\bigg\} \ + \ \tau_f^{(0,{\rm reg})}(T) \ \, ,
\label{tautreg}
\end{eqnarray}
where
\begin{equation}
\tau_f^{(0,{\rm reg})}(T) \ = \ 4 T \sum_{n=-\infty}^{\infty}
\sum_{c=r,g,b} \int \frac{ d^3 p}{(2\pi)^3} \
\frac{m_f}{\big(p_{nc}^2 + m_f^2\big)^2}
\ - 4 N_c \int \frac{ d^4 p}{(2\pi)^4} \
\frac{m_f}{\big(p^2 + m_f^2\big)^2} \ \, .
\label{t0treg}
\end{equation}
The expression in Eq.~(\ref{t0treg}) can be worked out, leading
to
\begin{eqnarray}
\tau_f^{(0,{\rm reg})}(T) & = &
- \frac{m_f}{T^3} \int \frac{ d^3 p}{(2\pi)^3} \
\bigg\{ \frac{1}{x_{f}^2} \bigg[
\frac{1}{1+\cosh x_{f}} \; + \;
\frac{8 + 4(3\Phi -1)\cosh x_{f}}{(3\Phi-1+2\cosh x_{f})^2}
\bigg] \nonumber \\
& & \ + \ \frac{6}{x_f^3} \; \frac{\Phi (1+2 e^{-x_f})
+e^{-2x_f}}{(1+e^{-x_f})\,({3\Phi-1+2\cosh x_{f}})} \bigg\} \ ,
\end{eqnarray}
where we have defined $x_f = \sqrt{p^2 + m_f^2}/T$. In the limit
of large temperature the behavior of the tensor coefficient is
given by
\begin{equation}
\tau_f(T)|_{T\to \infty} \ \simeq \ -\,
\frac{m_f\,N_c}{2\pi^2}\left[0.568 + \log(T/m_f)\right] \ .
\label{taularget}
\end{equation}

\section*{Appendix C: Tensor coefficient in the NJL model}

\renewcommand{\theequation}{C.\arabic{equation}}
\setcounter{equation}{0}

In this appendix we discuss the calculation of the tensor
coefficient in the (local) NJL model.
The value of $\tau_u$ that we have quoted in Table \ref{tableII} corresponds
to the SU(2) NJL model calculation carried out in Ref.~\cite{Frasca:2011zn}.
There, the authors use the Ritus formalism~\cite{Ritus:1972ky} to derive an
analytical expression for the VEV of the tensor polarization, and then they
introduce a smooth form factor in order to regularize the divergent momentum
integral. Our aim is to point out that there are alternative procedures that
can be followed to calculate $\tau_u$ within the NJL model. In fact, it is
seen that the numerical results turn out to be quite dependent on the way in
which the calculation is performed.

Let us briefly describe the procedure followed in
Ref.~\cite{Frasca:2011zn}. For consistency with this and other
previous calculations, throughout this Appendix expressions are
given in Minkowski space. The VEV of the tensor polarization
operator is given by
\begin{equation}
\langle\bar \psi_f\, \sigma_{\mu\nu}\, \psi_f\rangle_{A} \ = \
-\, i\, {\rm Tr}\left[ \sigma_{\mu\nu} \, S^{(A)}_f(x,x) \right]\ ,
\label{ten}
\end{equation}
where $S^{(A)}_f(x,x')$ is the $f$-quark propagator (in coordinate
space) in the presence of an external electromagnetic field
$A_\mu$. For the particular case of a constant magnetic field
$\vec B$ this propagator can be explicitly obtained. Within Ritus
formalism, choosing $\vec B$ to be along the $z$ axis, one has
\begin{equation}
S^{(A)}_f (x,x')_{\rm Rit} = \sum_{k=0}^\infty \frac{1}{(2 \pi)^4}
\int dp_0\, dp_2\, dp_3\ E_P(x) \; \Lambda_k\;
\frac{1}{\gamma\cdot P - M_f(B)}\; \bar E_P(x')\ ,
\label{ritus}
\end{equation}
where $E_P(x)$ stands for the eigenfunction of a charged fermion of momentum
$P^\mu$ in the magnetic field, and $\bar E_P(x) = \gamma_0 \,
E_P(x)^\dagger\, \gamma_0$. The index $k$ in the sum labels the Landau
levels (LL), while $\Lambda_k$ is a projector in Dirac space that takes into
account the LL degeneracy. The four-momentum $P^\mu$ is quantized according
to
\begin{equation}
P_\mu \ = \ \big( p_0\,,\,0\,,\, {\rm sign}(q_f) \sqrt{2 k |q_f| B}\,,\, p_3
\big)\ ,
\end{equation}
where $q_f$ denotes the quark electric charge. Replacing the
propagator in Eq.~(\ref{ritus}) into Eq.~(\ref{ten}), a
straightforward calculation shows that for our choice of magnetic
field orientation the 12 component of the tensor is the only one
that has a nonvanishing VEV. One has
\begin{equation}
\langle\bar \psi_f\, \sigma_{12}\, \psi_f\rangle_{A,{\rm Rit}} \ = \ N_c
\frac{q_f B}{2} \int \frac{dp_0\, dp_3}{(2 \pi)^2}\;
\frac{i}{p_0^2 - p_3^2 - M_f(B)^2}\ ,
\end{equation}
where the integral over $p_2$ has been performed. It is worth noticing that
only the lowest LL (i.e.~that corresponding to $k=0$) contributes to this
VEV. Now, this expression is divergent and needs to be regularized. In
Ref.~\cite{Frasca:2011zn} this has been achieved by introducing at this
stage a cutoff function $U_\Lambda(|\vec p|)$ that depends only on the
spatial components of the momentum, $\Lambda$ being a (three-momentum)
cutoff scale. Following this procedure and performing the integral over
$p_0$ one immediately obtains the expression in Eq.~(17) of
Ref.~\cite{Frasca:2011zn}. Expanding up to leading order in $B$, and noting
that $F^{12} = -B$, one gets within this method an explicit expression for
$\tau_f$, namely
\begin{equation}
\tau_f \ = \ \frac{N_c\, M_f}{2 \pi^2} \ I_{\rm Rit}(M_f/\Lambda) \ ,
\end{equation}
where
\begin{equation}
I_{\rm Rit}(x) \ = \ \int_0^\infty dy \
\frac{U_\Lambda\left(\Lambda y\right)}{\sqrt{y^2+x^2}}\ .
\label{ir}
\end{equation}
In Ref.~\cite{Frasca:2011zn} the particular form
\begin{eqnarray}
U^{\rm Lor5}_\Lambda( p ) \ = \ \frac{1}{1+(p/\Lambda)^{2 N}}\ , \qquad
N=5\ ,
\label{Lor5}
\end{eqnarray}
was used for numerical calculations. Alternatively, one can use a
simple sharp cutoff function $U^{\rm SC}_\Lambda\left( p \right)
= \theta\left(\Lambda-p\right)$, which leads to
\begin{eqnarray}
I^{\rm SC}_{\rm Rit}(x) \ = \ \log\left[ \frac{1 + \sqrt{1+x^2}}{x}
\right]\
\end{eqnarray}
(we have included the upper index SC to stress that it corresponds
to the particular case in which $U_\Lambda(p)$ is a sharp cutoff
function).

Let us consider now an alternative way to proceed based on the
so-called Schwinger proper-time representation of the fermion
propagator~\cite{Schwinger:1951nm}. As expected, once again it is
seen that only the VEV of the 12 component of the tensor is
nonvanishing. In this case one gets for this VEV the expression
\begin{equation}
\langle\bar \psi_f\, \sigma_{12}\, \psi_f\rangle_{A,{\rm Sch}} \ =
\ - N_c \frac{q_f B}{4\pi^2} \int_0^\infty \frac{ds}{s} \exp\left[-s
M_f^2(B)\right]\ ,
\end{equation}
which as in the previous case needs to be regularized. As is
customary when one uses the proper-time approach, we perform the
regularization by replacing the lower limit of the integral by
$1/\Lambda^2$. Expanding up to leading order in $B$ we get in this
way
\begin{equation}
\tau_f \ = \ \frac{N_c\, M_f}{2 \pi^2}
\, I_{\rm Sch}(M_f/\Lambda) \ ,
\end{equation}
where
\begin{equation}
I_{\rm Sch}(x) \ = \ \frac{1}{2} \, E_1(x) \ ,
\qquad
E_n(x)\ \equiv \ \int_1^\infty \; ds/s^n \, \exp(-s x)\ .
\end{equation}

Finally, a yet alternative way to proceed is to follow the steps discussed
in Sec.~II.A, in which we consider from the beginning the expansion of the
tensor operator at first order in powers of the magnetic field [see
Eq.~(\ref{tensor})], ending up with Eq.~(\ref{t0}). We refer to this
approach as ``weak field propagator expansion'' (WFPE). In the case of the
NJL model one has (in Minkowski space)
\begin{equation}
\langle\bar \psi_f\, \sigma_{\mu\nu}\, \psi_f\rangle_{A,{\rm WFPE}}
\ = \ -\,i\,4 N_c\, q_f\, F_{\mu\nu}
\int \frac{ d^4 p}{(2\pi)^4} \frac{M_f}{( p^2 - M_f^2)^2}\ .
\label{taunjl}
\end{equation}
It is important to stress that this expression can also be
obtained by considering the weak field limit of the fermion
propagator in a constant magnetic field given e.g.~in
Refs.~\cite{Chyi:1999fc,Watson:2013ghq}. As stated in
Ref.~\cite{Watson:2013ghq}, it is crucial to carry out the
infinite sum over Landau levels in order to obtain the proper form
of the propagator. Contrary to the case of the nonlocal model
discussed throughout Sect.~II, the integral in Eq.~(\ref{taunjl})
turns out to be divergent even in the chiral limit. By using a 3D
sharp cutoff regularization one finds
\begin{equation}
\tau_f \ = \ \frac{N_c M_f}{2 \pi^2} \ I^{\rm SC}_{\rm WFPE}
(M_f/\Lambda) \ ,
\end{equation}
where
\begin{equation}
I_{\rm WFPE}^{\rm SC}(x) \ = \ \log\left[ \frac{1+ \sqrt{1+x^2}}{x} \right]
-\frac{1}{\sqrt{1+x^2}}\ .
\end{equation}

We can now compare the results for $\tau_u$ within the SU(2) NJL
model that arise from the above discussed approaches. For this
purpose, in all cases we fix the model parameters in such a way
that the predicted values of $f_\pi$ and $m_\pi$ agree with the
corresponding empirical values and the light quark condensate
$\langle\bar \psi_u \, \psi_u\rangle^{1/3}$ has a
phenomenologically reasonable value of $-250$~MeV. Our numerical
results are given in Table~\ref{tabletau}.
\begin{table} [h]
\begin{center}
\begin{tabular}{ccccc}
\hline \hline
                  & Rit (Lor5)  & Rit (SC) &  Sch   &  WFPE (SC) \\
\hline
\hspace{.2cm} $\tau_u$   [MeV] \hspace{.2cm}
&  \hspace{.8cm}  70 \hspace{.8cm}    & \hspace{.8cm} 70  \hspace{.8cm}
&  \hspace{.8cm}  42 \hspace{.8cm}   & \hspace{.8cm}  28 \hspace{.8cm} \\
\hline \hline
\end{tabular}
\caption{Numerical values for the tensor coefficient $\tau_u$ in
the SU(2) NJL model}
\label{tabletau}
\end{center}
\end{table}
It is seen that our result for $\tau_u|_{\rm Rit}^{\rm Lor5}$ agrees with
the value given in Ref.~\cite{Frasca:2011zn}, which has been quoted in
Table~\ref{tableII}, and it is also coincident with the result obtained
within the Ritus approach for a sharp cutoff regularization function. On the
other hand, this value is significantly different from those obtained
following the Schwinger and WFPE approaches.
This shows that the results for the tensor coefficient obtained within the
local NJL model are quite dependent on the chosen regularization method, and
have to be taken with care. In order to understand the origin of this
dependence, it is interesting to compare with some detail the functions
$I(x)$ defined above. Since the cutoff $\Lambda$ is expected to be larger
than other scales in the problem, we can expand these functions for small
values of $x$. We get
\begin{eqnarray}
I_{\rm Rit}^{\rm LorN} &=& - \log x + \log 2 + \frac{\pi}{4 N} \csc \left(\frac{\pi}{N}\right) \ x^2 + {\cal O}(x^4)
\ ,
\nonumber \\
I_{\rm Rit}^{\rm SC}   &=& - \log x + \log 2 + \frac{1}{4} \ x^2 + {\cal O}(x^4)
\ ,
\nonumber \\
I_{\rm Sch}        &=& - \log x - \frac{\gamma}{2} + \frac{1}{2} \ x^2 + {\cal O}(x^4)
\ ,
\nonumber \\
I_{\rm WFPE}^{\rm SC}   &=& - \log x + (\log 2 - 1) + \frac{3}{4} \ x^2 + {\cal
O}(x^4) \ ,
\label{diff}
\end{eqnarray}
where $\gamma$ is the Euler constant. As expected, all expressions have the
same leading logarithmic contribution in the limit of small $x$, therefore
the corresponding predictions for the tensor coefficient will be similar for
$M_f/\Lambda \ll 1$. However, for realistic values of this ratio, the
constant terms and the contributions carrying powers of $x$ become relevant.
The differences between these terms shown in Eqs.~(\ref{diff}) explain the
differences between the numerical results in Table~\ref{tabletau}.

\end{document}